\documentclass[aps,prb,twocolumn,amssymb,showpacs,superscriptaddress]{revtex4}
\usepackage{bm}
\usepackage{graphicx}
\begin{document}
\title{Impurity effects in a two--dimensional system with Dirac spectrum}
\author{Yu.V.~Skrypnyk}
\affiliation{G.~V.~Kurdyumov Institute of Metal Physics,
             National Academy of Sciences of Ukraine,
             Vernadskogo Avenue 36, Kyiv 03142, Ukraine.}
\author{V.~M.~Loktev}
\affiliation{Bogolyubov Institute for Theoretical Physics,
             National Academy of Sciences of Ukraine,
             Metrolohichna Street 14-b, Kyiv 03143, Ukraine}

\begin{abstract}
It is demonstrated that in a two--band 2D system the resonance state is manifested
close to the energy of the Dirac point in the electron spectrum for the sufficiently
large impurity perturbation. With increasing the impurity concentration, the electron
spectrum undergoes the rearrangement, which is characterized by the opening of the
broad quasi--gap in the vicinity of the nodal point. If the critical concentration
for the spectrum rearrangement is not reached, the domain of localized states remains
exponentially small compared to the bandwidth.
\end{abstract}


\maketitle

The effect of impurities on quasi--particle spectra in disordered systems is
qualitatively determined by the ratio of the dimensionality of the disordered system
to the exponent in the respective dispersion relation. The disordered system exhibits
low--dimensional behavior when this ratio is less than unity. In general, impurity
effects are more pronounced in low--dimensional systems. Materials like graphene are
certainly 2D objects\cite{gr1,gr2}. However, electrons in graphene feature the linear
dispersion close to the Fermi level. A number of experiments evidently demonstrate that
graphene is highly tolerant to impurity induced perturbations. This fact can be
attributed to the increased effective dimensionality of the electron subsystem in
graphene. With respect to the ordinary quadratic dispersion graphene could be regarded
as a four--dimensional system. This high effective dimensionality should be beneficial
for the reduction of localization effects that occur due to impurities, which are
inevitably present (or intentionally introduced) in corresponding materials.

The importance of impurity effects for the physics of graphene had been frequently
emphasized. Notwithstanding, the effect of disorder were studied only in both weak
scattering\cite{Khv1} and unitary\cite{vitor,per} limits, or for a kind of
interpolation between such extreme cases\cite{Khv2}. When impurity states of single
defects are located in the vicinity of the van Hove singularities of the host system,
an increase in the impurity concentration yields a substantial spectrum rearrangement
(SR), albeit the relative impurity concentration remains quite low\cite{crosrear,lif}.
This transition between two qualitatively different regimes of impurity scattering
takes place only for a finite magnitude of the single--impurity perturbation. The type
of the state that is produced by the single impurity is usually reflected in the
passage of the SR. Below we are attempting to examine a possibility for impurity states
to appear close to the Dirac point of the electron spectrum in a 2D system with linear
dispersion for an arbitrary strength (unitary limit including) of the single--impurity
perturbation, and to outline a scenario of the SR with varying the impurity
concentration. Similar issues have been raised up in Ref.~\onlinecite{pog} but the
problem have not been solved correctly.

In order to model a system with the Dirac spectrum, one can choose the host
tight--binding Hamiltonian in the most basic form\cite{ham1}, 
\begin{equation}
\hat{H}_{0}=\sum_{\bm{k}}[f(\bm{k}) c^{\dag}_{1}(\bm{k}) c^{}_{2}(\bm{k})+
f^{*}(\bm{k}) c^{\dag}_{2}(\bm{k}) c^{}_{1}(\bm{k})], \label{h0}
\end{equation}
where $c^{\dag}_{\alpha}(\bm{k})$ and $c^{}_{\alpha}(\bm{k}))$ are creation and
annihilation operators on two sublattices, and $\bm{k}$ is a 2D wave vector. Since
only the close vicinity of the nodal point will be of concern, it is sufficient to put
$f(\bm{k})=t a (k_{x}+i k_{y})$, $t>0$, where $t$ is the hopping parameter, and $a$ is
the lattice constant. Then, the dispersion relation, $\epsilon(\bm{k})=\pm t a k$,
does possess a Dirac point at the zero energy, which separates two bands that are
touching each other.

We also assume that our system can be reasonably well described as a substitutional
binary alloy with a diagonal disorder (a so-called Lifshits model). It is supposed that
impurities are distributed absolutely at random on both sublattices, so that on--site
potentials can take one of two values, say $V_{L}$ and $0$, with probabilities $c$ and
$1-c$, respectively. The full Hamiltonian of the disordered system is then represented
by the sum of the translationally invariant host part (\ref{h0}) and the
perturbation,
\begin{equation}
\hat{H} = \hat{H}_0 + \frac{V_{L}}{N}\sum_{\bm{k},\bm{k'},<\alpha,p>}%
e^{i(\bm{k'}-\bm{k})\bm{r}_p}c^{\dag}_{\alpha}(\bm{k})c^{}_{\alpha}(\bm{k'}),
\label{fullH}
\end{equation}
where $<\alpha,p>$ ranges over those sites on the lattice that are occupied by
impurities. 

\begin{figure}
\includegraphics[width=0.475\textwidth,clip]{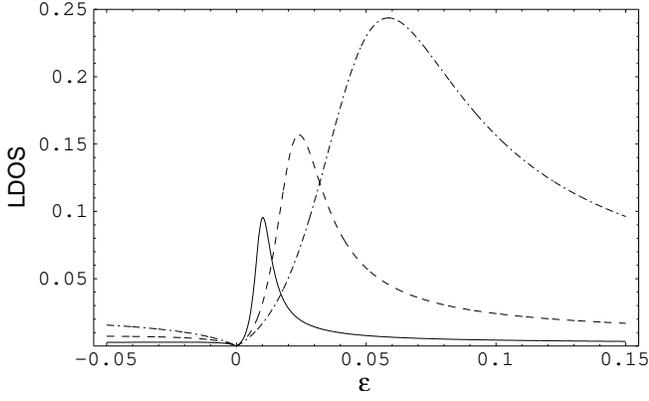}
\caption{\label{f1} LDOS at the impurity site for $v=-10,-5,-2.5$ is shown by solid,
dashed and dot--dashed curves, respectively.}
\end{figure}

Let only the zeroth site on one of sublattices be occupied by an impurity. Then, the
diagonal element of the Green's function (GF) $\hat G=(\epsilon-\hat H)^{-1}$ on this
site
\begin{equation}
G_{0}=g_{0}/(1-V_{L}g_{0}), \label{gf}
\end{equation}
where $g_{0}$ is the diagonal element of the GF in the host,
$\hat g=(\epsilon-\hat H_{0})^{-1}$. Site--diagonal elements $g_{0}$ are equal on
both sublattices and can be easily obtained by approximating the Brillouin zone
with a circle,
\begin{eqnarray}
g_{0}&=&\frac{1}{N}\sum_{\bm{k}}\frac{\epsilon}{\epsilon^{2}-\epsilon(\bm{k})^{2}}%
=\frac{a^{2}}{2 \pi}\int_{0}^{2\sqrt{\pi}/a}\frac{\epsilon k\, dk}%
{\epsilon^{2}-t^{2} a^{2} k^{2}}= \nonumber\\
&=&\frac{\epsilon}{4\pi t^{2}}\ln\left(\frac{\epsilon^{2}}{4\pi t^{2}-%
\epsilon^{2}}\right)-i\frac{\left|\epsilon\right|}{4t^{2}},\;%
\left|\epsilon\right| \leqslant 2\sqrt{\pi}t. \label{g0}
\end{eqnarray}
It is convenient to choose the energy unit in such a way that the bandwidth
$2\sqrt{\pi}t=1$. Thus, for (\ref{g0}) one obtains
\begin{equation}
g_{0}=\epsilon\ln(\epsilon^{2}/(1-\epsilon^{2}))-%
i\pi\left|\epsilon\right|. \label{gd}
\end{equation}
The local density of states (LDOS) at the impurity site (see Fig.~\ref{f1}) is given
by the imaginary part of the diagonal element of the GF (\ref{gf}),
\begin{equation}
\rho_{0}=-\frac{1}{\pi}\Im G_{0}=\frac{\left|\epsilon\right|}%
{(1-v\epsilon\ln\left[\epsilon^{2}/(1-\epsilon^{2})\right])^{2}+(v\pi\epsilon)^{2}},
\label{ldos}
\end{equation}
where $v$ is the dimensionless single--impurity perturbation. For the sufficiently
large $|v|$, a prominent peak is manifested in the LDOS (\ref{ldos}) close to the
Dirac point in the spectrum, indicating the presence of the resonance state. Its
energy $\epsilon_r$ is defined by the Lifshits equation
\begin{equation}
1\approx 2v\epsilon_{r}\ln\left|\epsilon_{r}\right|.
\end{equation}  
It should be emphasized that for the attractive impurity potential $v<0$ the
energy $\epsilon_r$ is located above the nodal point ($\epsilon_r>0$), and, vice versa,
it is located below this point ($\epsilon_r<0$) for the repulsive impurity potential
$v>0$. In contract to 3D systems, the resonance state is accompanied by the deep local
level outside of both adjacent bands. Thus, the total number of states near the nodal
point is gradually diminishing with increasing $|v|$. 

When the resonance peak is relatively narrow, the denominator in Eq.~(\ref{ldos})
can be expanded about $\epsilon_r$,
\begin{eqnarray}
\rho_{0}&\approx&\frac{\left|\epsilon\right|\Gamma^{2}}{(v\pi\epsilon_{r})^{2}%
\left[(\epsilon-\epsilon_{r})^{2}+\Gamma^{2}\right]},\nonumber \\
\Gamma&=&\pi\left|\epsilon_{r}\right|\left|\ln(\epsilon_{r}^{2}/%
(1-\epsilon_{r}^{2}))+2/(1-\epsilon_{r}^{2})\right|^{-1}.\label{gam}
\end{eqnarray}
The resonance state is well-defined when the effective damping (\ref{gam}) is much less
than its separation from the closest van Hove singularity,
\begin{equation}
\Gamma/\left|\epsilon_{r}\right|\approx\pi/%
[2\left|\ln\left|\epsilon_{r}\right|+1\right|]\ll 1.
\end{equation} 
The inequality is satisfied only when the resonance energy $\epsilon_{r}$ is located
fairly close to the Dirac point and is strengthening with decreasing $|\epsilon_{r}|$.
Thus, the resonance presence in the unitary limit\cite{vitor,per} is justified. It is
worth mentioning that well-defined resonances can not appear in the vicinity of the
band edge in a single--band 2D or 3D system within the Lifshits model (\ref{fullH}).

It is not difficult to calculate also the change $\Delta\rho$ in the total DOS in the
system that is caused by the single impurity center\cite{lif},
\begin{eqnarray}
&&\Delta\rho=\frac{v}{N\pi}\Im\bigl(\frac{dg_{0}}{d\epsilon}%
\frac{1}{1-vg_{0}}\bigr)= \nonumber \\
&&=\frac{-v\mathop{\mathrm{sign}}(\epsilon)(1-\epsilon^{2}+2v\epsilon)}%
{N\Bigl(1-\epsilon^{2}\Bigr)\Bigl\{\Bigl[1-v\epsilon\ln\left(\frac{\epsilon^{2}}%
{1-\epsilon^{2}}\right)\Bigr]^{2}+\Bigl[v\pi\epsilon\Bigr]^{2}%
\Bigr\} }.
\end{eqnarray}
It can be verified that bare states are redistributed within the bands. For the case
$v<0$, states are removed from the domain of the continuous spectrum in the lower band
to the split off local level, and in the upper band states are pushed towards
$\epsilon_r$. 
However, there is a notable negative dip in $\Delta\rho$ at the nodal point, where the
host DOS is zero (see Eq.~(\ref{gd})). Therefore, the close vicinity of the nodal point
can not be described properly by the direct expansion in the impurity concentration
even at the negligibly small concentration of impurities. 

Commonly, renormalized methods, such as the co\-he\-rent--potential approximation 
(CPA), are the most effective inside the continuous spectrum. The one--electron GF of
the disordered system can be expressed by the corresponding self--energy
$\hat{\Sigma}(\bm{k})$. Since the translational invariance is restored by the
configurational averaging over impurity distributions,
\begin{equation}
\hat{G}(\bm{k})^{-1}=\hat{g}(\bm{k})^{-1}-\hat{\Sigma}(\bm{k}),
\end{equation} 
where operators $\hat{G}(\bm{k})$, $\hat{g}(\bm{k})$, and $\hat{\Sigma}(\bm{k})$ are
acting in the sublattice space. For the model system under consideration, the
self--energy within the CPA is site--diagonal and identical on both sublattices.
According to the conventional procedure, it should be determined in a self--consistent
manner from the equation,
\begin{equation}
\sigma=c v/(1-(v-\sigma)g_{0}(\epsilon-\sigma)). \label{cpa}
\end{equation}
In the effective medium constructed by the CPA, the self--energy can be expanded into
the series in impurity clusters\cite{crosrear,lif},
\begin{equation}
\Sigma^{\alpha\beta}(\bm{k})=\delta^{\alpha\beta}\sigma+\sigma^{\alpha\beta}_{2}%
(\bm{k})+\dots, \label{ser}
\end{equation}
where $\hat{\sigma}_{2}(\bm{k})$ represents the contribution from pair diagrams,
\begin{eqnarray}
&\sigma^{\alpha\beta}_{2}&\!\!\!\!(\bm{k})=\delta^{\alpha\beta}%
\sum_{m,n,\,l\gamma\ne0\alpha}\xi_{m}\xi_{n}\frac{\tau^{3}_{m}\tau^{2}_{n}%
\bigl(\mathfrak{G}^{\alpha\gamma}_{0l}\bigr)^{2}%
\bigl(\mathfrak{G}^{\gamma\alpha}_{l0}\bigr)^{2}}%
{1-\tau_{m}\tau_{n}\mathfrak{G}^{\alpha\gamma}_{0l}%
\mathfrak{G}^{\gamma\alpha}_{l0}}+  \nonumber \\
&+&\sum_{m,n,\,l\beta\ne0\alpha}\xi_{m}\xi_{n}\frac{\tau^{2}_{m}\tau^{2}_{n}%
\bigl(\mathfrak{G}^{\alpha\beta}_{0l}\bigr)^{2}\mathfrak{G}^{\beta\alpha}_{l0}%
\exp(i\bm{k}\bm{r}_l)}{1-\tau_{m}\tau_{n}%
\mathfrak{G}^{\alpha\beta}_{0l}\mathfrak{G}^{\beta\alpha}_{l0}}. \label{s2}
\end{eqnarray}   
In Eq.~(\ref{s2}) $l$ and indices $\alpha$, $\beta$, and $\gamma$ enumerate lattice
cells and sublattices, respectively,
\begin{equation}
\mathfrak{G}^{\alpha\beta}_{0l}=\frac{1}{N}\sum_{\bm{k}}g^{\alpha\beta}%
(\epsilon-\sigma,\bm{k})\exp(-i\bm{k}\bm{r}_l),
\end{equation}
the single--site T--matrix is denoted by
\begin{equation}
\tau_{m}=(\upsilon_{m}-\sigma)/[1-(\upsilon_{m}-\sigma)g_{0}(\epsilon-\sigma)],
\end{equation}
while indices $m$ and $n$ enumerate atom types (impurity or host), so that $\xi_{m}$
attains values $c$ or $1-c$ depending on the value of these indices, and variable
$\upsilon_{m}$ is $v$ or $0$, respectively.

The relative magnitude of contributions from scatterings on impurity clusters is
increasing as approaching any van Hove singularity in the spectrum, so that the CPA
becomes unreliable in their vicinity. The necessity to implement a relevant
applicability criterion for the CPA and other approximate methods based on the partial
summation of the series for the GF have been overlooked in some recent articles
devoted to the impurity effects in graphene\cite{per}. The analysis of the series
expansion for $\hat{\Sigma}(\bm{k})$ shows that the series does have a small parameter,
\begin{equation}
R(\epsilon)=\sum_{m}\xi_{m}(\tau_{m})^{2}%
\sum_{l\beta\ne0\alpha}\bigl(\mathfrak{G}^{\alpha\beta}_{0l}\bigr)^{2}.\label{small}
\end{equation}
Cluster diagrams can be omitted on $|R(\epsilon)|\leqslant 1/2$. Inside the energy
domains, where this inequality holds, only the first term can be retained in the series
and the resulting approximate expression for the self--energy does not depend on
$\bm{k}$. If the relative impurity concentration is kept low, multiple--occupancy
corrections that are included in the derivation of the CPA can be neglected too, so it
is reduced to the so--called method of modified propagator,
\begin{equation}
\sigma=cv/(1-vg_{0}(\epsilon-\sigma)).\label{mp}
\end{equation}  
Since our interest is restricted to the narrow vicinity of the nodal point in the
spectrum, it is possible to make an obvious approximation for the diagonal element
of the host GF,
\begin{equation}
g_{0}\approx2\epsilon\ln\left|\epsilon\right|-i\pi\left|\epsilon\right|,\,%
\left|\epsilon\right|\ll 1. 
\end{equation}
By making a substitution $\epsilon-\sigma=\varkappa \exp(i\varphi)$, $0<\varphi<\pi$,
the imaginary part of Eq.~(\ref{mp}) can be rewritten as follows,
\begin{eqnarray}
&&cv^{2}\left[2\ln\varkappa+(2\varphi-\pi)\cot\varphi\right]+ \nonumber\\
&+&\left[1-v\varkappa(2\ln\varkappa\cos\varphi-(2\varphi-\pi)\sin\varphi\right)]%
^{2}+ \nonumber \\
&+&\left[v\varkappa(2\ln\varkappa\sin\varphi+(2\varphi-\pi)\cos\varphi\right)]^{2}%
=0. \label{sim} 
\end{eqnarray}
Starting from some threshold magnitude of $\varkappa$, there are two solutions of
Eq.~(\ref{sim}) for the phase $\varphi$ at the given concentration of impurities,
which correspond to the two existing bands. Respective values of $\epsilon$ are then
provided by the real part of Eq.~(\ref{mp}), which closes up the parametric solution of
the problem. Correspondingly, the validity criterion for the CPA assumes the form,
\begin{equation}
|R(\epsilon)|\approx\left|\frac{\ln\varkappa+1+i(\varphi-\frac{\pi}{2})}%
{\ln\varkappa+(\varphi-\frac{\pi}{2})\cot\varphi}\right|\leqslant\frac{1}{2}.
\label{appl}
\end{equation}  
 
As usual, for the renormalized wave vector in both bands one has
$\tilde{k}a=2\sqrt{\pi}\varkappa\left|\cos\varphi\right|$. The spatial behavior of the
host GF on one of sublattices at large intercell distances is given by
\begin{eqnarray}
g_{0l}&=&\frac{a^{2}}{(2 \pi)^{2}}\int_{0}^{2\sqrt{\pi}/a}%
\!\!\!\!\frac{\epsilon k\, dk}{\epsilon^{2}-(a k)^{2}/(4\pi)}\int_{0}^{2\pi}%
e^{ikr_{l}\cos\phi}\,d\phi\approx \nonumber \\
&\approx&2\epsilon\int_{0}^{\infty}\frac{J_{0}(ur_{l}/a)u\, du}%
{4\pi\epsilon^{2}-u^{2}}=\pi\epsilon[Y_{0}(2\sqrt{\pi}|\epsilon|r_{l}/a)-%
\nonumber \\
&-&i\mathop{\mathrm{sign}}(\epsilon)J_{0}(2\sqrt{\pi}|\epsilon|r_{l}/a)],\label{g0l}
\end{eqnarray}
where $J_{0}$ and $Y_{0}$ are the Bessel functions of the 1st and 2nd kind,
respectively. It follows from Eq.~(\ref{g0l}) that the mean free path should be written
as $\ell=a/(4\sqrt{\pi}\varkappa\sin\varphi)$. Thus, the localization parameter from
the Ioffe-Regel criterion\cite{ioffe} takes the simple form,
$\tilde{k}\ell=|\cot\varphi|/2$.

An overview of the SR scenario can be provided based on simple estimations. It may seem
that $\epsilon=\sigma=cv$ is an appropriate solution of Eq.~(\ref{mp}). However, this
is not the case. Formally, this equation is satisfied, but an analytical solution for
the GF that is passing through this point can not be constructed. On the other hand,
there should be always an energy, at which $\Re(\epsilon-\sigma)=0$. As follows from
Eq.~(\ref{sim}), a certain amount of damping ($\Im\sigma\ne0$) is always present at
this energy,
\begin{equation}
-2cv^{2}\ln\varkappa_{0}=1+(2v\varkappa_{0}\ln\varkappa_{0})^{2}. \label{fin}
\end{equation} 
When the impurity concentration is sufficiently small,
$\varkappa_{0}\approx\exp(-1/(2cv^{2}))$. It is not difficult to see from the real part
of Eq.~(\ref{mp}) that $\Re\sigma\approx cv$ in this case. In other words, the energy,
at which two bands coincide, is shifted approximately by $cv$ from zero towards the
impurity local level. The width of the concentration smearing area around
$\epsilon\approx cv$, where states are highly localized according to the Ioffe--Regel
criterion, should be proportional to $\varkappa_{0}$, while the guess value for the
mean free path inside this area remains exponentially large. For small $\varkappa$,
Eq.~(\ref{sim}) is reduced to
\begin{equation}
cv^{2}(\ln\varkappa+(\pi-2\varphi)\cot\varphi)\approx-1.
\end{equation}
In the same approximation, it follows from the real part of Eq.~(\ref{mp}) that
\begin{equation}
\epsilon-cv\approx cv^{2}(\pi-2\varphi)\varkappa/\sin\varphi.
\end{equation} 
Although the threshold magnitude of the localization parameter, which separates states
that can be described by the wave vector, can be argued to a known extent, it seems
reasonable to choose it from the thoroughly tested method of potential--well
analogy\cite{econ1,econ2,econ3}, $|\cot\varphi|>\sqrt{3}$. Then, the width of the
concentration smearing area is
\begin{equation}
\Delta_{IR}\approx(8\pi/3)\exp(\pi/\sqrt{3})cv^{2}\exp(-1/(2cv^{2})).\label{dir}
\end{equation} 
In the narrow vicinity of $\epsilon\approx cv$, contributions from scatterings on
impurity clusters are becoming significant. According to the applicability
criterion Eq.~(\ref{appl}), the electron spectrum obtained by the CPA can not be
justified inside the area with the width of $\Delta_{R}\approx\exp(-1/(4cv^{2}))/e$,
which is wider than $\Delta_{IR}$ that follows from the Ioffe-Regel criterion. It had
been shown that in the 3D systems the small parameter of the series expansion
Eq.~(\ref{small}) and the localization parameter $\tilde{k}\ell$ can be expressed
through each other\cite{former} and depend on the phase $\varphi$ only. However, in the
system under consideration the cut--off phase for the CPA applicability criterion
depends on the disorder parameter $cv^{2}$ at the small impurity concentration. The
reason of this discrepancy is the subject of the more detailed study.   

With an increase in the impurity concentration, the absolute value of the shift
$|cv|$ and the width of the concentration smearing area are also gradually increasing
in magnitude. It is obvious from  the expression for $\varkappa_{0}$, which can be
rewritten as $\varkappa_{0}\approx -c|v|(2|v|\varkappa_{0}\ln\varkappa_{0})$ that
parametrically $|cv|$ and $\varkappa_{0}$ simultaneously become of the order of the
resonance energy $|\epsilon_{r}|$. The second area of concentration smearing opens in
the vicinity of the resonance energy and, finally, both areas of concentration
smearing are merged together. This is indicative of the spectrum rearrangement.
Both criteria are coinciding in this regime ($\Delta_{R}\approx\Delta_{IR}$). An
expression for the critical concentration of the SR can be obtained by comparing two
main parameters of the problem by their magnitude,
\begin{equation}
cv^{2}\exp[-1/(2cv^{2})]=\zeta c|v|,
\end{equation}
where $\zeta$ is a certain constant to be determined. This immediately yields
\begin{equation}
c_{r}=-1/(2v^{2}\ln(\zeta/|v|)).
\end{equation}
This expression fits well calculated critical concentrations of the SR with
$\zeta\sim 10^{-5}$.
 
At the impurity concentration that is far exceeding this critical value
(i.e. $c\gg c_{r}$),
it follows from Eq.~(\ref{fin}) that in the first approximation $\varkappa_{0}$ does
not depend on the magnitude of the impurity perturbation $v$ . Both criteria give
similar
results for the width of the broad concentration smearing area, $\Delta_{R}\sim%
\Delta_{IR}\sim\sqrt{-(c)/\ln(\sqrt{c})}$, which is nearly symmetric about the Dirac
point of the host system. As was mentioned above, analogous approach to the
description of the impurity effects in graphene have been undertaken in
Ref.~\onlinecite{pog}, in which some miscalculations were committed in course
of the theoretical analysis of the problem. Nevertheless, the width of the
concentration smearing area for $c\gg c_{r}$ have been estimated correctly.

While the passage of the SR in the system with linear dispersion deserves more close
attention, some conclusions can be made at this stage. When the change in the on--site
potential caused by the impurity atom is noticeably larger than the bandwidth, the
well--defined resonance state can do appear in the system with the Dirac spectrum.
However, this resonance is not very sharp for the reasonable amount of the impurity
perturbation. As a rule, the presence of the well--defined resonance state leads to the
SR of the cross type with an increase in the impurity concentration. Yet, there are
some exceptions from this rule, and the system under consideration belongs to
them\cite{jap}. Despite the resonance, the SR is of the anomalous type that is common
in low--dimensional systems. This anomalous SR is characterized by the opening of the
quasi--gap, in which any adequate cluster expansion can not be constructed and states
are highly localized. The electron spectrum is not much distorted outside of the
concentration smearing area, and there are no prominent features in it close to the
resonance energy. 

When the change in the on--site potential on the impurity site is not extremely large,
the SR does not occur at all, and the width of the quasi-gap remains exponentially
small. Indeed, from the practical point of view, such exponentially small quasi--gap
will remain unnoticed in most situations, and virtually does not affect the carrier
mobility. In the case of the large change in the on--cite potential disorder effects
are not significant until the critical concentration of the impurities is reached. The
obtained results also apply to the systems with the gap in the host quasi--particle
spectrum when this gap is less than the width of the concentration smearing area.

We are deeply grateful to Prof. V.~P.~Gusynin, who turned our attention to the problem
of impurity states in systems with Dirac spectrum.


\begin{thebibliography}{99}

\bibitem{gr1}M.~S.~Dresselhaus and G.~Dresselhaus, Adv. Phys. {\bf 51}, 1 (2002).

\bibitem{gr2}K.~S.~Novoselov{\it et al.}, Nature {\bf 438}, 197 (2005).

\bibitem{Khv1} D.~V.~Khveshchenko, cond-mat/0602398.

\bibitem{vitor}V.~M.~Pereira {\it et al.}, Phys. Rev. Lett. {\bf 96} 036801 (2006).

\bibitem{per}N.~M.~R.~Peres, F.~Guinea, A.~H.~Castro Neto, Phys. Rev. B {\bf 73},
125411 (2006).

\bibitem{Khv2}D.~V.~Khveshchenko, cond--mat/0604180.

\bibitem{crosrear}M.~A.~Ivanov, Sov. Phys. Solid State {\bf 12}, 1508
(1971).

\bibitem{lif}I.~M.~Lifshits, S.~A.~Gredeskul, and L.~A.~Pastur, 
{\it Introduction to the Theory of Disordered Systems} (Wiley, N.~Y., 1988).

\bibitem{pog}Yu.~G.~Pogorelov, cond--mat/0603327.

\bibitem{ham1}G.~W.~Semenoff, Phys. Rev. Lett. {\bf 53}, 2449 (1984).

\bibitem{ioffe}A.~F.~Ioffe and A.~R.~Regel, Prog. Semicond. 4, 237
(1960).

\bibitem{econ1}C.~M.~Soukoulis {\it et al.}, Phys. Rev. Lett. {\bf 62}, 575 (1989).

\bibitem{econ2}E.~N.~Economou and C.~M.~Soukoulis, Phys. Rev. B {\bf 40}, 7977 
(1989).

\bibitem{econ3}A.~Kirchner, K.~Busch, and C.~M.~Soukoulis, Phys. Rev. B {\bf 57},
277 (1998).

\bibitem{former} Yu.~Skrypnyk, Phys. Rev. B {\bf 70}, 212201 (2004).



\bibitem{jap}Yu.~V.~Skrypnyk and B.~I.~Min, Progr. Theor. Phys. {\bf 108}, 1021 (2002).

\end{thebibliography}
\end{document}